\documentclass{aa}  

\usepackage{graphicx}
\usepackage{txfonts}
\usepackage{amsmath}
\usepackage{xspace}
\usepackage{multirow}
\usepackage[section]{placeins}
\usepackage{overpic}
\usepackage{subcaption}
\usepackage[normalem]{ulem}
\usepackage{hyperref}
\usepackage{xcolor}
\usepackage{hyperref}
\usepackage{float}
\makeatletter
\renewcommand*\aa@pageof{, page \thepage{} of \pageref*{LastPage}}
\makeatother
\hypersetup{colorlinks=true,allcolors=[rgb]{0,0,0.8}}

\begin{document}

   \title{Non-halo structures and their effects on gravitationally lensed galaxies}

   \author{Baptiste Jego
          \inst{1,2,3},
          Giulia Despali
          \inst{4,5,6}
          Tamara Richardson
          \inst{7}
          \and
          Jens Stücker
          \inst{8}
          }

   \institute{Université de Strasbourg, CNRS, Observatoire astronomique de Strasbourg (ObAS), UMR 7550, F-67000 Strasbourg, France\\
             \email{baptiste.jego@astro.unistra.fr}
              \and
              ENS Paris-Saclay, 4 Av. des Sciences, 91190 Gif-sur-Yvette, France
              \and
              Max Planck Institute for Astrophysics, Karl-Schwarzschild-Strasse 1, D-85748 Garching bei München, Germany
              \and
              Dipartimento di Fisica e Astronomia "Augusto Righi", Alma Mater Studiorum Università di Bologna, via Gobetti 93/2, I-40129 Bologna, Italy
              \and
              INAF-Osservatorio di Astrofisica e Scienza dello Spazio di Bologna, Via Piero Gobetti 93/3, I-40129 Bologna, Italy
              \and
              INFN-Sezione di Bologna, Viale Berti Pichat 6/2, I-40127 Bologna, Italy
              \and
              Donostia International Physics Center (DIPC), Paseo Manuel de Lardizabal, 4, 20018, Donostia-San Sebasti\'an, Gipuzkoa, Spain
              \and
              Institute for Astronomy, University of Vienna, Türkenschanzstraße 17, Vienna 1180, Austria
             }

    \date{Received YYY; accepted ZZZ}
 
  \abstract
   {
   While the $\Lambda$CDM model is very successful on large scales, its validity on smaller scales remains uncertain. Recent works suggest that non-halo dark matter structures, such as filaments and walls, could significantly influence gravitational lensing and that the importance of these effects depends on the dark matter model: in warm dark matter scenarios, fewer low-mass objects form and thus their mass is redistributed into the cosmic-web. We investigate these effects on galaxy-galaxy lensing using fragmentation-free Warm Dark Matter (WDM) simulations with particle masses of m$_{\chi}$ = 1 keV and m$_{\chi}$ = 3 keV. Although these cosmological scenarios are already observationally excluded, the fraction of mass falling outside of haloes grows with the thermal velocity of the dark matter particles, which allows for the search for first-order effects. We create mock datasets, based on gravitationally-lensed systems from the BELLS-Gallery, incorporating non-halo contributions from these simulations to study their impact in comparison to mocks where the lens has a smooth mass distribution.
   Using Bayesian modelling, we find that perturbations from WDM non-halo structures produce an effect on the inferred parameters of the main lens and shift the reconstructed source position. However, these variations are subtle and are effectively absorbed by standard elliptical power-law lens models, making them challenging to distinguish from intrinsic lensing features. Most importantly, non-halo perturbations do not appear as a strong external shear term, which is commonly used in gravitational lensing analyses to represent large-scale perturbations. Our results demonstrate that while non-halo structures can affect the lensing analysis, the overall impact remains indistinguishable from variations of the main lens in colder WDM and CDM scenarios, where non-halo contributions are smaller.
   }

   \authorrunning{Baptiste Jego et al}
   
   \maketitle

\section{Introduction}\label{sec:intro}

    One of the long-lasting problems and main goals of modern cosmology is to understand the nature of dark matter \citep[see e.g.][]{Bosma1981, Bullock_2017, galaxies7040081}. In the Lambda Cold Dark Matter ($\Lambda$CDM) paradigm, often referred to as the standard model of cosmology, dark matter particles are assumed to be non-relativistic, cold, and to make up for $85$ per cent of the matter content and approximately $27$ per cent of the total mass-energy budget of the Universe (see \cite{2020A&A...641A...6P}). $\Lambda$CDM is widely used in numerical simulations, as it can accurately reproduce and predict the large-scale structure of the Universe \citep{2012AnP...524..507F, SFW}, and provides a physical explanation for several crucial phenomena such as the CMB (Cosmic Microwave Background) \citep{2011A&A...536A..18P}, the large-scale distribution of galaxies, and the accelerated expansion of the Universe.

    Several models have been proposed as alternatives to CDM \citep[see e.g.][for reviews]{Bullock_2017,Bertone_2018} to reconcile predictions from $\Lambda$CDM with observations on galactic and sub-galactic scales. Here, we will focus on models that suppress small-scale structure formation, specifically thermal relic Warm Dark Matter (WDM) models \citep{Bode_2001}. In these models, dark matter particles are produced in an equilibrium state with a non-negligible initial thermal velocity. This initial thermal velocity dispersion allows the dark matter particles to free-stream out of the gravitational potential wells of smaller perturbations and, as a result, suppresses the gravitational collapse of these perturbations.

    Strong gravitational lensing allows one to measure the dark matter distribution on the sub-galactic scale where the difference between models is large. Hence, it provides a robust avenue for testing predictions from different dark matter models \citep{Dalal_2002, 10.1111/j.1365-2966.2009.15230.x, Vegetti_2018, 2020MNRAS.491.6077G, 10.1093/mnrasl/slad074} through the detection of low-mass dark-matter structures, of which the abundance is highly sensitive to the nature of dark matter. Regarding the search for thermal relic dark matter, this approach is also complementary to other methods such as the analysis of stellar streams \citep{Banik_2021}, the Lyman-$\alpha$ forest \citep{Ir_i__2024} and Milky Way satellites \citep{PhysRevLett.126.091101} with which it can be combined to produce stronger constraints \citep{10.1093/mnras/stab1960}. However, a careful understanding of all possible sources of systematic errors is needed. For example, \cite{10.1093}, hereafter cited as R22, investigate the impact of non-halo structures on flux-ratio anomalies in multiply imaged quasars. By sampling 1000 random lines of sight, the authors show that in a $m_{\chi} = 3$ keV WDM scenario, neglecting the non-halo material can lead to an underestimation of the flux ratio anomalies by five to ten per cent, which increases substantially for the warmer 1 keV model. From this analysis, the authors conclude that material outside of haloes, such as filaments and walls, should be included in rigorous models and that their inclusion can affect the constraints on the DM particle mass, leading to a potential bias in favour of colder dark matter models when neglected.
    
    In this paper, we investigate whether these non-halo structures also affect the surface brightness distribution of strongly lensed arcs. The latter is sensitive to local changes of the first derivative of the lensing potential in a way that is strongly dependent on the angular resolution of the data: the better the resolution, the smaller the angular size of the perturbations that can be detected \citep{Despali_2019}. Here, we aim to study the impact that non-halo structures have on lensed arc observations in different WDM scenarios. To this effect, we generate and analyse a set of mock images reproducing observations of two systems which have been studied as part of the BELLS-Gallery sample \citep{Shu_2016_2, 10.1093/mnras/stad3694, 10.1093/mnras/sty2833}.
    We use the new, fragmentation-free simulations from R22 to add non-halo structures to mock lensing datasets and test if they can be detected and distinguished from the effect of the main lens. 
    Recent studies on quadruply lensed quasars have ruled out DM models with $m_{\chi} <5.2$ keV, favouring colder DM particles \citep{10.1093/mnras/stz3177,10.1093/mnras/stz3480}. However, these analyses neglect the effect of material outside of haloes. Analysis of the WDM non-halo structures could relax these constraints. However, because these structures are extremely difficult to model, we are limited to the parameters used in R22 and their numerical requirements. As such, we consider two models with WDM particle masses $m_{\chi} = 1$ and  $m_{\chi} = 3$ keV. We note that, although $m_{\chi} = 1$ keV is ruled out by observations, in this work this model serves as an interesting limiting case, allowing us to gauge the amplitude of the perturbations.
    
    The paper is structured as follows: in Sect.~\ref{sec:sims} we present the data from our WDM simulation and for the gravitational imaging. In Sect.~\ref{sec:data} we detail the mock data which we model in Sect.~\ref{sec:model}. Then in Sect.~\ref{sec:results} we present and discuss our results and present our conclusions in Sect.~\ref{sec:conc}.

\section{Simulations}\label{sec:sims}

    \begin{figure}
        \centering
        \includegraphics[width=\hsize]{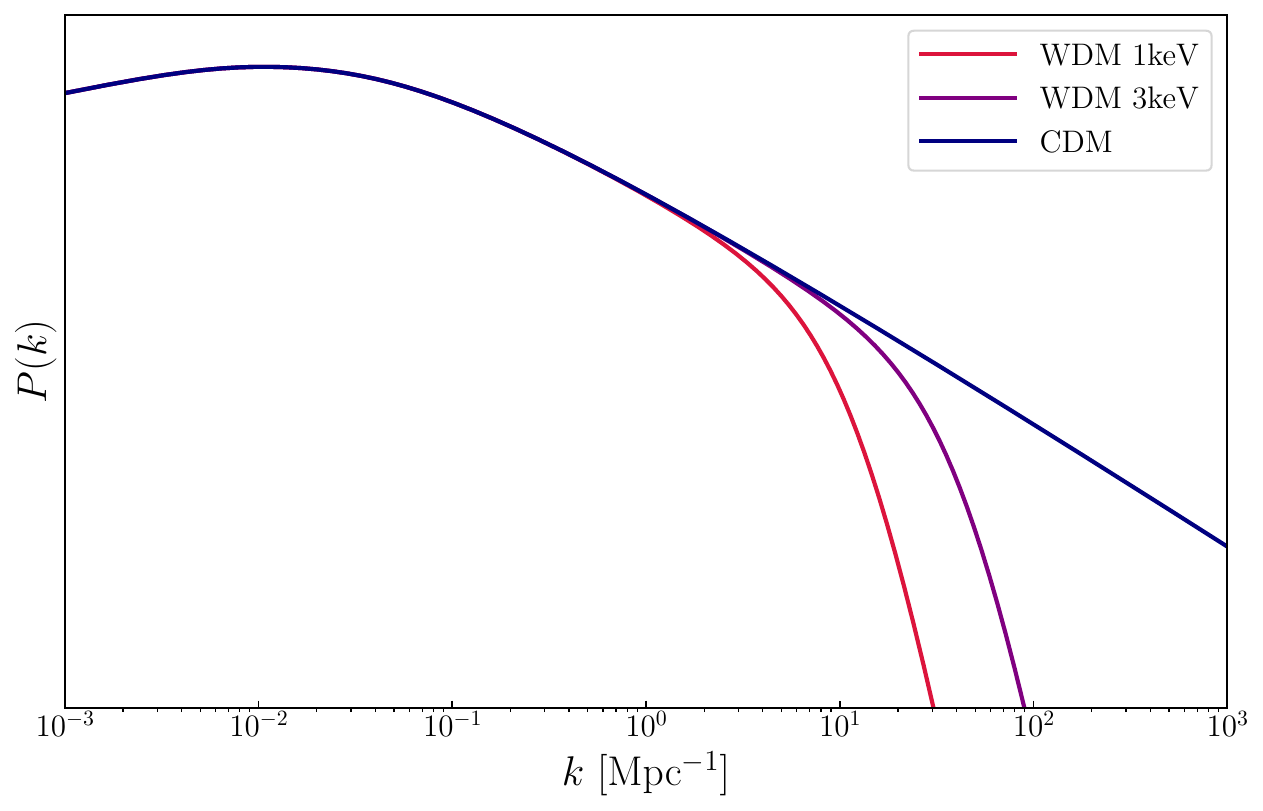}
        \caption{Power spectra for CDM, $m_{\chi} = 1$ keV and $m_{\chi} = 3$ keV WDM. The WDM power spectra correspond to low-pass filtered versions of the CDM power spectrum. The vertical axis is unitless and arbitrarily normalised. The free-streaming scale is a function of the dark matter particle mass $m_{\chi}$, and is directly related to the cut-off scale.}
    \label{fig:spectra}
    \end{figure}
      
    \begin{figure*}
        \centering
        \includegraphics[width=\textwidth]{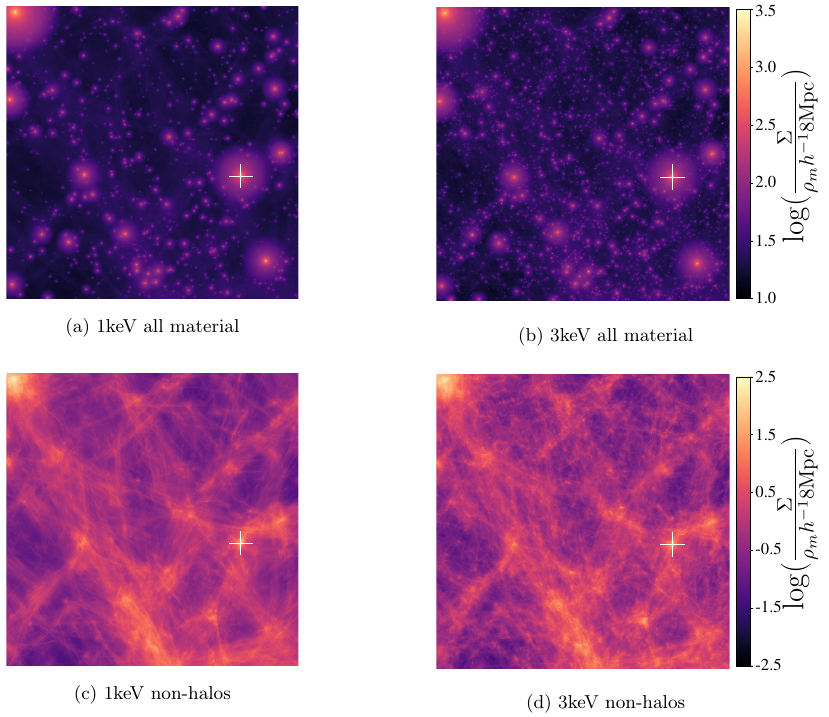}
        \caption{Upper panel: Full projected data from both simulations from R22. Each square has a side length of $8h^{-1}$Mpc, and corresponds to a  $80h^{-1}$Mpc projection depth. Panel (b) contains fewer small-scale dark matter haloes than panel (a) as the free-streaming length is longer in 1 keV than in 3 keV WDM. Lower panel: Non-Halo projected data from both simulations from R22. Each square has a side length of $8h^{-1}$Mpc$^{1}$, and corresponds to a  $80h^{-1}$Mpc$^{1}$ projection. Panel (c) contains smoother and less clumpy structures at small scales than panel (d). The markers are all at the same coordinates and show the regions from which we select the non-halo material used as additional perturbations in our gravitationally lensed systems. These correspond to the highest density of non-halo material in both simulations and to the densest halo in both simulations, excluding those which are cut in the top-left corner.}
    \label{fig:diff_1_3}
    \end{figure*}

    \begin{figure*}
        \centering
        \includegraphics[width=\textwidth]{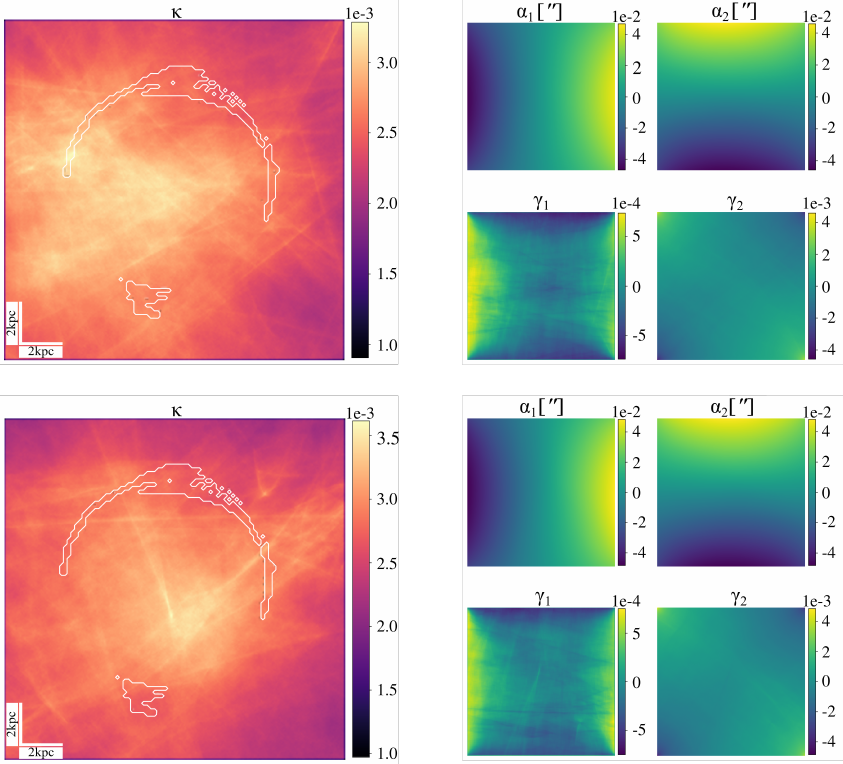}\caption{$20\times20$ kpc$^{2}$ maps of the convergence, deflection angle components, and shear components for the non-halo material in two regions around haloes at the same coordinates in the 3 keV (top) and 1 keV (bottom) simulations from R22. The deflection angle maps are used to add the effect of non-halo structures on top of the deflection from the elliptical power law lenses in our datasets. The positions of the non-halo material from which we compute these convergence maps are marked in Fig.~\ref{fig:diff_1_3}. They are the same in both simulations. The white contours are based on the arc light image of SDSSJ110+3649 propagated from $z=0.733$.}
    \label{fig:maps}
    \end{figure*}

\subsection{Simulation parameters and models}\label{ssec:psims_params}

    In this work, we analyse two runs from the suite of WDM simulations presented in \citet{2022MNRAS.509.1703S}. 
    Each volume represents a $(20h^{-1}{\rm Mpc})^3$ periodic box, run with cosmological WDM initial conditions generated using MUSIC\footnote{\href{https://www-n.oca.eu/ohahn/MUSIC/}{https://www-n.oca.eu/ohahn/MUSIC/}} \citep{2013ascl.soft11011H}. Both simulations were run using the same initial Gaussian random noise field and with the same cosmological parameters, $h = 0.679$, $\Omega_{\rm m} = 0.3051$, $\Omega_{\Lambda} = 0.6949$, $\Omega_{\rm K} = 0$, and $\sigma_{8} = 0.8154$ \citep{2020A&A...641A...6P}, but differ with regards to the dark matter model they represent. In particular, we use two volumes replicating cosmological structure formation in the presence of thermal relic dark matter with a particle mass of either, $m_{\chi} = 1{\rm keV}$ or $m_{\chi} = 3{\rm keV}$, which we will hereafter refer to as the 1 keV or 3 keV simulations respectively.

    In WDM cosmologies the thermal velocity of DM particles leads them to have a finite free-streaming length \citep{Bode_2001},
    \begin{equation}
        \lambda_{\rm s} = \frac{0.048}{h\text{Mpc}^{-1}}\left(\frac{\Omega_{\chi}}{0.4}\right)^{0.15}\left(\frac{h}{0.65}\right)^{1.3}\left(\frac{1\text{keV}}{m_{\chi}}\right)^{1.15}\left(\frac{1.5}{g_{\chi}}\right)^{0.29},
    \label{eq:Ls}
    \end{equation}
    which depends on the particle mass $m_{\chi}$, the dark matter density in units of the critical density of the Universe, $\Omega_{\chi}$, and $g_{\chi}$, a dimensionless factor which accounts for the number of degrees of freedom of the dark matter particles, here we adopt $g_{\chi} = 1.5$, the value corresponding to the case of sterile neutrinos.

    This free-streaming motion is considered negligible in CDM, but in the case of WDM dampens the power spectrum on small scales. Here, this dampening is modelled using the \citet{Bode_2001} transfer function,
    \begin{equation}
        P_{\rm WDM}(\textbf{k}) = (1 + (\lambda_{s} k)^{2})^{-10} P_{\rm CDM}(\textbf{k}),
    \label{eq:spectra}
    \end{equation}
    where $P(\textbf{k})$ is the scale-dependant power spectrum, and $\textbf{k}$ is the wavenumber vector. 
    The resulting dampening to the power spectrum is shown in Fig.~\ref{fig:spectra}.

    Beyond smoothing out density fluctuations, the free-streaming motion of the WDM particles allows them to escape small-scale gravitational potential fluctuations, which effectively suppresses structure formation below the half-mode mass scale,
    \begin{equation}
        M_{\rm hm} = \frac{4\pi}{3}\rho_{\rm m}\left(\frac{\lambda_{\rm hm}}{2}\right)^{3}.
    \label{eq:Mhm}
    \end{equation}
    where $\rho_{\rm m}$ is the mean matter density of the Universe, and
    \begin{equation}
        \lambda_{\rm hm} = 2\pi\lambda_{\rm s}(2^{1/5} - 1)^{-1/2}.
    \label{eq:Lhm}
    \end{equation}
    $M_{\rm hm}$ corresponds to the mass at which the WDM power spectrum is suppressed by a factor of two compared to its CDM counterpart \citep{Viel_2005, Schneider_2012}. The WDM models used here respectively result in a half-mode mass of 
    $M_{hm}=2.5\times10^{10}h^{-1}$M$_{\odot}$ for the 1 keV model and $M_{hm}=5.7\times10^{8}h^{-1}$M$_{\odot}$ for the 3 keV model. 
    
    WDM models are commonly associated with the lack of low-mass structures. While this is the main feature of these models, it is worth noting that the material which is not confined to haloes remains within non-halo structures, for instance, R22 find that, at $z=0$, the fraction of matter outside of haloes represents $f_{\rm non-halo} = 45.7$ per cent of the total dark matter mass in the 1 keV model, and respectively $f_{\rm non-halo} = 34.8$ per cent in the 3 keV model. This means that in WDM models, more matter is redistributed in the smooth cosmic web large-scale structures compared to CDM, where $f_{\rm non-halo}$ is around $5\% - 20\%$ and increases with redshift (e.g. \cite{10.1111/j.1365-2966.2009.15742.x} for the 100GeV neutralino CDM).

\subsection{Avoiding artificial fragmentation}\label{ssec:sims_frag}
    
    It has been found that in $N$-body WDM simulations the smoother density field will artificially fragment due to discreteness effects \citep{10.1111/j.1365-2966.2007.12053.x} resulting in a large population of spurious small mass haloes. Although methods have been developed to identify and remove this spurious population \citep[see e.g.][]{lovell14,2022MNRAS.509.1703S}, these artificial haloes have nevertheless prohibited the study of the smoother regions of the density fields in this type of simulation.

    To circumvent this issue, another approach has been developed \citep{Abel_2012, PhysRevD.85.083005, 2013ascl.soft11011H} which relies on the assumption that the DM distribution function only occupies a three dimensional hyper-surface of the full six-dimensional phase space. In these simulations, the phase0space distribution function is tessellated using a set of tracer particles, which can in turn be used to calculate the temporal evolution of the full phase space distribution function with much higher accuracy. However, these methods break down inside virialised structures due to chaotic mixing, requiring an exponentially large number of tracer particles to accurately trace these systems \citep{2016MNRAS.455.1115H, 2016JCoPh.321..644S}.

    The simulations used here make use of a hybrid tessellation-$N$-body method \citep{St_cker_2020,2022MNRAS.509.1703S}, which dynamically separates particles into four categories: voids, walls, filaments, and haloes. For the first three categories, the phase space distribution function is tessellated using $N_{\rm V} = 256^3$ particles, while the final category is modelled using $N_{\rm T} = 512^3$ tracer particles of mass $m_{\rm T} = 5.0\cdot10^6h^{-1}{\rm M}_\odot$ which are released in regions where the density field has collapsed along three dimensions, i.e. inside haloes. This hybrid approach maintains the efficiency of $N$-body inside haloes while allowing the large-scale distribution function to be traced with very high accuracy. 

    An additional advantage of the hybrid tessellation-$N$-body approach is that it allows, in post-processing, to effectively interpolate the smooth density field down to arbitrarily small masses. allowing the creation of extremely high-resolution density maps outside of haloes. It is using this property that R22 were able to project ten $(8\times8\times8)h^{-3}{\rm Mpc}^3$ subvolumes with an effective mass resolution of $\sim 20h^{-1}{\rm M}_\odot$, which, once put together, represent a continuous $(80\times8\times8)h^{-3}{\rm Mpc}^3$ projection. These projected density maps are shown in Fig.~\ref{fig:diff_1_3}: in the upper panels, we show all the matter inside the simulation, except the halos that have been fitted and replaced by spherical NFW profiles to reduce the numerical noise; in the lower panels, we only show the matter present in non-halo structures, showing the 1 keV and 3 keV simulations in the left and right panels, respectively. As discussed previously, we can see that the 3 keV simulation contains many more low-mass haloes and that the 1 keV simulation appears far smoother with fewer visible clumps in the filaments and walls.
    
    \begin{table}
        \caption{Parameters for the WDM cosmological simulations used in R22 and in this work.}
        \label{tab:Sim}
        \centering
        \begin{tabular}{c c c}
            \hline\hline
            Parameter & 1 keV & 3 keV \\ 
            \hline
            $h$ & 0.679 & - \\
            $\Omega_{\rm m}$ & 0.3051 & - \\
            $\Omega_{\Lambda}$ & 0.6949 & - \\
            $\Omega_{\rm K}$ & 0 & - \\
            $\sigma_{8}$ & 0.8154 & - \\
            $N_{\rm V}$ & 256$^{3}$ & - \\
            $N_{\rm T}$ & 512$^{3}$ & - \\
            $m_{\rm T}$ $[h^{-1}$M$_{\odot}]$ & $5.0\cdot 10^{6}$ & - \\
            $f_{\rm non-halo}$ & 45.7\% & 34.8\% \\
            $M_{\rm hm}$ $[h^{-1}$M$_{\odot}]$& $2.5\cdot10^{10}$ & $5.7\cdot10^{8}$ \\
            $\lambda_{\rm s}$ [kpc] & 5.06 & 1.44 \\   
            $\lambda_{\rm hm}$ [kpc] & 82.6 & 23.4 \\
            \hline
        \end{tabular}
        \tablefoot{The cosmological parameters are the best-fit values from the Planck collaboration 2018 results \citep{2020A&A...641A...6P}.}
    \end{table}

\subsection{Lensing maps}\label{ssec:sims_maps}

    From the simulated non-halo surface density maps (lower panels in Fig.~\ref{fig:diff_1_3}), we compute lensing properties using the convention of \cite{2001PhR...340..291B}. We define the convergence $\kappa$ as the ratio between the surface mass density and a critical surface density which is a function of the lens and the source angular diameters:
    \begin{equation}
        \kappa \equiv \frac{\Sigma(\boldsymbol{x})}{\Sigma_{\rm cr}}, \text{ with } \Sigma_{\rm cr} = \frac{c^{2}}{4\pi G}\frac{D_{\rm S}}{D_{\rm L}D_{\rm LS}},
    \label{eq:kappa}
    \end{equation}
    with $c$ the speed of light, $G$ the gravitational constant, and $D_{\rm L}$, $D_{\rm S}$, and $D_{\rm LS}$ respectively being the angular diameter distance from the observer to the lens, the observer to the source, and from the lens to the source.

    From the convergence, we compute the lensing potential $\Psi$ using the Poisson equation $\Delta_{x}\Psi(\boldsymbol{x}) = 2\kappa$. In turn, from the lensing potential we define the deflection angle maps, $\alpha_{1}$ and $\alpha_{2}$, as the individual components of the gradient, $\nabla_{x}\Psi(\boldsymbol{x}) = \boldsymbol{\alpha}(\boldsymbol{x})$, along with the shear components $\gamma_{1}$ and $\gamma_{2}$ as
    \begin{equation}
        \gamma_{1}(\boldsymbol{x}) \equiv \frac{1}{2}(\Psi_{11} - \Psi_{22}),
    \label{eq:gamma1}
    \end{equation}
    and
    \begin{equation}
        \gamma_{2}(\boldsymbol{x}) \equiv \Psi_{12} = \Psi_{21},
    \label{eq:gamma2}
    \end{equation}
    where $\Psi_{ij} = \frac{\partial^{2} \Psi}{\partial x_{i}\partial x_{j}}$.

    In Fig.~\ref{fig:maps} we provide a set of example fields produced from two $20\times20$ kpc$^{2}$ cut-outs of the projected density field of the same region as seen in both simulations. No direct comparison between the 3 keV and 1 keV maps can be made because the non-halo structures display different shapes and characteristics even though they are situated at the same coordinates. Nonetheless, we observe that all the quantities have systematically higher values for 1 keV than for 3 keV which is consistent with the higher fraction of matter outside of haloes, $f_{\rm non-halo}$, in the warmer model - see the values of $f_{\rm non-halo}$ given in Tab.~\ref{tab:Sim}.

    A key parameter to consider when extracting the effect of the non-halo material from the simulations is the projection depth. Here, due to computational limitations, we only have access to data with a maximum depth of 80$h^{-1}$Mpc, while typical lensed sources are located at high redshifts, resulting in observer-source distance of the order of a few Gpc. 
    However, $\Sigma_{\rm cr}$ peaks at the typical lens redshift value for a certain source redshift, which implies that the line-of-sight perturbing material lying far away from the main lens, such as low-mass field haloes, has a negligible effect on the lensed system \citep{10.1093/mnras/sty159}. The dependence of the relative impact of the non-halo material with respect to haloes on this projection depth is studied and quantified in the case of quadruply-lensed quasars in R22, where the authors show that the improvements in the analysis are effectively minor when considering longer lines of sight as they study relative contributions of the non-halo structures to small-mass halos. In this work, we also study the relative contributions of non-halo structures to unperturbed lens models. Therefore, because we have the same computational limitations as R22, and do not look at absolute effects, we choose to make use of the full 80$h^{-1}$Mpc depth which is enough to quantify the impact of the non-halo structures while avoiding repeating the same structures too many times. Thus, we use the same projection size as R22.

\section{Mock data}\label{sec:data}

    \begin{figure*}
        \centering
        \includegraphics[width=\textwidth]{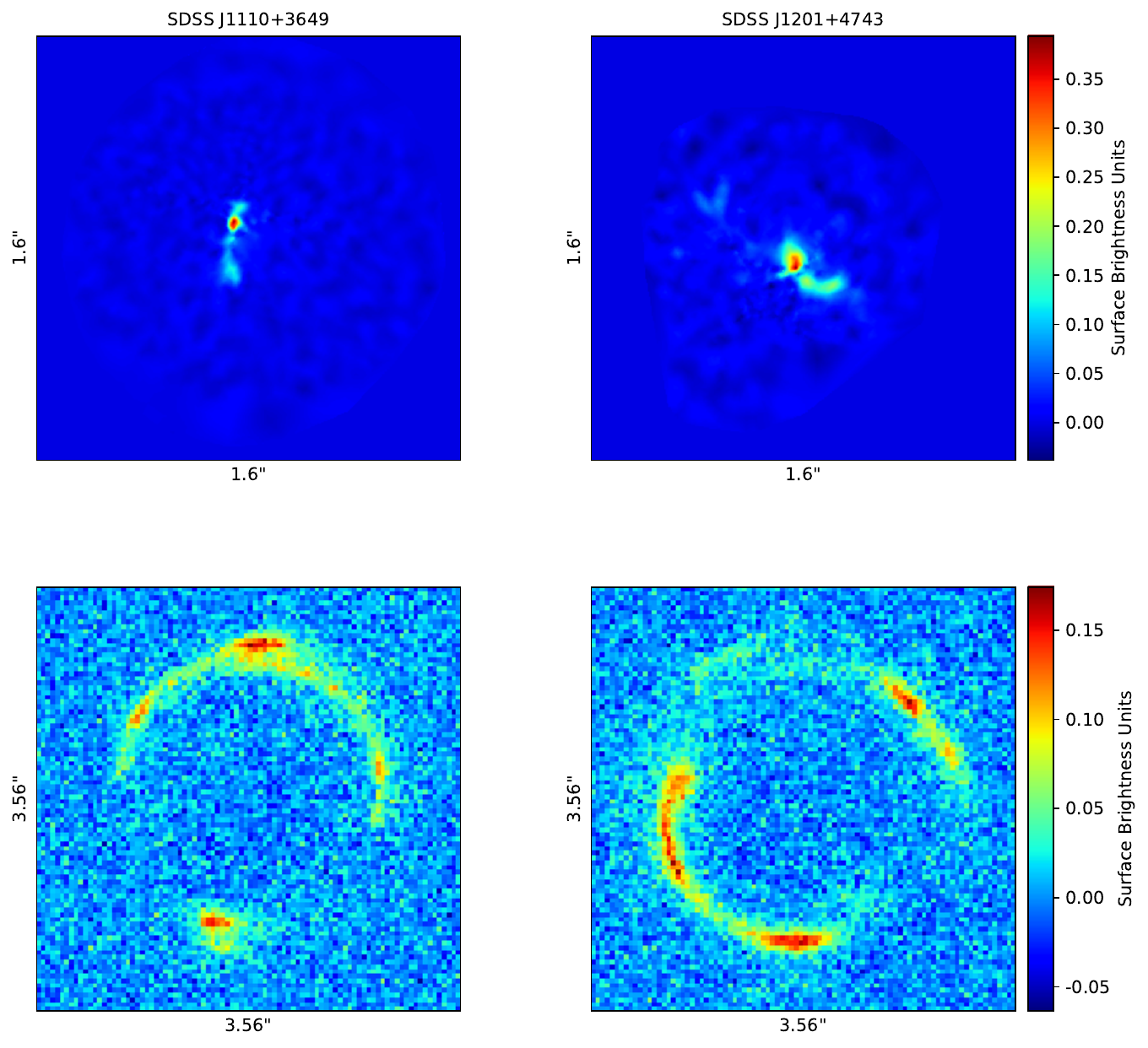}
        \caption{The left panels represent the system SDSS J1110+3649, and the right panels represent the system SDSS J1201+4743. The top panels represent the source brightness distribution, and the bottom panels represent the observed flux after propagating through the lens, convolving with an HST-type point spread function, and adding a Gaussian noise resulting in a signal to noise ratio of 10.}
    \label{fig:systems}
    \end{figure*}
    
    We created mock lensing observations based on two strong gravitational lens systems from the eBOSS Emission-Line Lens Survey for GALaxy-Ly$\alpha$ EmitteR sYstems (BELLS GALLERY, \cite{Shu_2016}): SDSSJ1110+3649 and SDSSJ1201+4743. We choose these specific systems because they have been studied in detail in previous works and have a reasonably good signal-to-noise level \citep{Despali_2019}. They were observed with the WFC3-UVIS camera and the F606W filter on the Hubble Space Telescope. These systems have been independently modelled by \cite{Shu_2016_2}, \cite{10.1093/mnras/stad3694} and \cite{10.1093/mnras/sty2833}, and we use the lens model parameters from the latter. Observational details on the base systems are given in Tab.~\ref{tab:obs}. Technical details are also provided in this table, where the Einstein radii are from \cite{Ritondale_2019}. We created mock observations that reproduce the original HST observations in terms of resolution and noise by convolving the obtained map with an HST-type point-spread function (PSF) and adding a random noise field. We generate the noise field by independently sampling a normal random variate with a standard deviation ten times lower than the maximum brightness of the image, as to obtain a signal-to-noise ratio, $\frac{S}{N}=10$, in agreement with the typical uncertainties on the grade-A lenses magnitudes from the BELLS-Gallery \citep{2016ApJ...833..264S}. The noise maps for the unperturbed and perturbed maps of the same system are computed independently. These two systems are presented in Fig.~\ref{fig:systems} with the respective image and source.

    \begin{table}
        \caption{Summary of the image acquisition characteristics, and source and lens redshifts of both gravitationally lensed systems studied in this work.}
        \label{tab:obs}
        \centering
        \begin{tabular}{c c c c}
            \hline\hline
            Characteristic & SDSSJ1110+3649 & SDSSJ1201+4743 \\
            \hline
            Instrument & HST/WFC3-UVIS & - \\
            Filter & F606W & - \\
            Exp time [s] & 2540 & 2624 \\
            Field of view ["$^{2}$] & 3.56$\times$3.56 & - \\
            Pixel size & 90$\times$90 & - \\
            $z_{\rm l}$ & 0.733 & 0.563 \\
            $z_{\rm s}$ & 2.502 & 2.126 \\
            $\lambda_{\rm r}$ [\mbox{\normalfont\AA}] & 1682 & 1883 \\
            $R_{\rm e}$ ["] & 1.04 ± 0.02 & 1.72 ± 0.04 \\
            &  & 0.94 ± 0.03 \\
            \hline
        \end{tabular}
        \tablefoot{$z_{s}$ and $z_{l}$ are the source and lens redshifts, $\lambda_{r}$ is the rest-frame UV emission, and $R_{e}$ is the Einstein radius. SDSSJ1201+4743 has two Einstein radii as the lens has two components.}
    \end{table}

    From these systems, we created three datasets: $\mathcal{D}_{1}$ and $\mathcal{D}_{2}$ based on SDSSJ1110+3649 and $\mathcal{D}_{3}$, based on SDSSJ1201+4743. Each dataset contains three variations: 
    \begin{itemize}
        \item a mock image created from the lens parameters and mock source with no additional perturbation (”Unperturbed”),
        \item a mock image where perturbations in the lens plane from simulated 1 keV WDM filaments and walls (”1 keV”) are added to the smooth case,
        \item an image with perturbations in the lens plane from simulated 3 keV WDM filaments and walls (”3 keV”).
    \end{itemize}
    
    We include the WDM perturbations to the deflection angles through the convergence maps described in the previous section. These are computed from a subregion of the full projected non-halo density maps. This subregion, marked by the white cross in Fig.~\ref{fig:diff_1_3}, is chosen to have the highest density and thus maximises the perturbative effect of the non-halo structures. Although this does not allow for a statistical study, we perform a quantitative estimation of the effect of the non-halo structures and make sure that our conclusions are not affected by the specific geometry or alignment effects in the chosen regions. This is explained in more detail in Sect.~\ref{sec:results}. 
    
    For both systems, we model the main deflectors using the elliptical power-law lens parameters from \cite{Ritondale_2019}. We add the lensing effect computed from the non-halo structure density fields to these elliptical smooth lenses. It is through the sum of these two deflectors, main lens and non-halo perturbations, that we propagate the source surface brightness distribution to produce the mock observed flux distribution. The perturbations from the simulated non-halo 1 keV and 3 keV dark matter structures on the left column of Fig.~\ref{fig:maps} are the same for $\mathcal{D}_{1}$ and $\mathcal{D}_{3}$, but have been rotated by 90 degrees for $\mathcal{D}_{2}$ to check that our results are not biased by a particular alignment or anti-alignment of the filaments/walls with intrinsic geometries of the lens mass model. Similarly, $\mathcal{D}_{3}$ is used to check that the results obtained with $\mathcal{D}_{1}$ are representative of typical gravitational lensing systems rather than outliers.

\section{Lens modelling}\label{sec:model}
    
    To quantify the effect of non-halo structures, we model the mock data sets described above using the \textsc{pronto} code (Vegetti et al. in prep), a grid-based Bayesian modelling technique developed by \citet{10.1111/j.1365-2966.2008.14005.x}, and further extended by  \citet{10.1093/mnras/sty2594}, \citet{Ritondale_2019}, \citet{10.1093/mnras/staa2740} and \citet{ndiritu2024selfconsistentframeworkstudymagnetic}. We simultaneously infer the most probable parameters of the lens mass density distribution, $\boldsymbol{\eta_{m}}$, along with the brightness distribution and smoothness of the background source. While the latter is modelled pixel by pixel in a regularised way, the former is modelled using an elliptical power-law profile \citep{1975ApJ...195...13B, 10.1093/mnras/208.3.511, 1994A&A...284..285K} given by

    \begin{equation}
        \kappa(x,y) = \frac{\kappa_{0}(2-\frac{\gamma}{2})q^{\gamma-3/2}}{2(q^{2}(x^{2}+r_{c}^{2})+y^{2})^{(\gamma-1)/2}},
    \label{eq:lensmass}
    \end{equation}

    where $\kappa_{0}$ is the surface mass density normalisation, $q$ is the axis ratio, $\gamma$ is the radial slope of the mass-density profile, and $r_{c}$ is the core radius, which is not a free parameter and is kept at $10^{-4}$ arcseconds.
    Two additional parameters account for external shear: the shear strength $\Gamma$ and the shear angle $\Gamma_{\theta}$.
    Finally, multipoles can be added as an extension to the elliptical power law + external shear profile. Multipoles of order $m$ create a convergence $\kappa_{m}$ of the form

    \begin{equation}
        \kappa_{m}(r,\phi) = r^{1-\gamma}[a_{m}\sin(m\phi) + b_{m}\cos(m\phi)],
    \label{eq:mtp}
    \end{equation}

    where r and $\phi$ are the polar coordinates in the reference frame of the lens. These multipoles introduce asymmetries in the lens geometry, and one can measure their ability to absorb the perturbations due to external mass components \citep{10.1093/mnras/stae153, 10.1093/mnras/stac2350}. Here, we will use multipoles of the $3^{rd}$ order or of the $3^{rd}$ and $4^{th}$ orders. Therefore, the most extensive form of the free parameter vector is $\boldsymbol{\eta_{m}} = (\kappa_{0}, q, \gamma, x_{c}, y_{c}, \theta, \Gamma, \Gamma_{\theta}, a_{3}, b_{3}, a_{4}, b_{4})$ with $x_{c}$, $y_{c}$, and $\theta$ the coordinates of the centre and the orientation of the lens. The values of the fiducial parameters used to create the mock data and the prior probability distributions are given in Tab.~\ref{tab:mock_123}.
    
    \begin{table}
        \caption{Values of the lens mass distribution fiducial parameters used to create the mock data and prior probability distributions to model it for $\mathcal{D}_{1}$, $\mathcal{D}_{2}$ and $\mathcal{D}_{3}$.}
        \label{tab:mock_123}
        \centering
        \begin{tabular}{c c c c c}
            \hline\hline
            Param. & $\mathcal{D}_{1}$, $\mathcal{D}_{2}$ & $\mathcal{D}_{3}$ & Prior($\mathcal{D}_{1}$, $\mathcal{D}_{2}$) & Prior ($\mathcal{D}_{3}$)\\ [0.5ex] 
            \hline
            $\kappa_{0}$ & 1.14 & 1.19 & $\mathcal{U}[1.0,1.4]$ & -\\
            $q$ & 0.82 & 0.78 & $\mathcal{U}[0,1]$ & -\\
            $\gamma$ & 0.51 & 0.47 & $\mathcal{U}[0,1]$ & -\\
            $x_{c}$ & 0.02 & -0.13 & $\mathcal{U}[0.0,0.2]$ & $\mathcal{U}[-0.2,0]$\\
            $y_{c}$ & 0.15 & -0.18 & $\mathcal{U}[0.0,0.2]$ & $\mathcal{U}[-0.2,0]$\\
            $\theta$ & 80.0 & 38.5 & $\mathcal{U}[50,100]$ & $\mathcal{U}[0,50]$\\
            $\Gamma$ & 0 & -0.01 & $\mathcal{U}[-0.05,0.05]$ & -\\
            $\Gamma_{\theta}$ & 0 & 41.2 & $\mathcal{U}[-20,20]$ & $\mathcal{U}[-10,50]$\\
            $a_{3}$ & 0 & 0 & $\mathcal{U}[-0.02,0.02]$ & -\\
            $b_{3}$ & 0 & 0 & $\mathcal{U}[-0.02,0.02]$ & -\\
            $a_{4}$ & 0 & 0 & $\mathcal{U}[-0.02,0.02]$ & - \\
            $b_{4}$ & 0 & 0 & $\mathcal{U}[-0.02,0.02]$ & - \\
            \hline
        \end{tabular}
        \tablefoot{All the priors are uniform and cover a reasonably large range of values around the fiducial value.}
    \end{table}

    \begin{table}
        \caption{Values of the differences in logarithmic evidence from the nested sampling of the parameters of each model of lens mass distribution for $\mathcal{D}_{1}$.}
        \label{tab:Ev_D1}
        \centering
        \begin{tabular}{c c c c}
            \hline\hline
            $\Delta$log($\mathcal{E}$) & Unperturbed & 1 keV & 3 keV \\ [0.5ex] 
            \hline
            Reference & $\Delta_{0}$ & $\Delta_{0}$ & $\Delta_{0}$ \\
            Fixed $\Gamma_{\theta}$,$\Gamma$ & -0.3±0.3 & -4.0±0.3 & -1.5±0.3 \\
            Multipoles $m=3$ & -21.5±0.3 & -15.6±0.3 & -34.3±0.3 \\
            Multipoles $m=3, 4$ & -26.5±0.3 & -21.5±0.3 & -37.2±0.3 \\
            \hline
        \end{tabular}
        \tablefoot{The results are given in the cases of the Unperturbed system, and the system with additional perturbations from the 1 keV and 3 keV simulations non-halo material. In each column, we compare three of our models with the reference one, where $\Delta_{0}$ is the reference logarithmic evidence value, and negative values indicate lower evidence for the more complex models.}
    \end{table}

    We explore the parameter space with the nested sampling algorithm \texttt{MultiNest} \citep{Feroz_2009}. To compare two different models of the same dataset, we compare the logarithmic value of the evidence, log($\mathcal{E}$). A lower value of log($\mathcal{E}$) corresponds to a better fit of the model to the data, and therefore the difference $\Delta$log($\mathcal{E}$) = log($\mathcal{E}$) - $\Delta_{0}$, where $\Delta_{0}$ is a reference value, is a quantitative way to compare the different models used for our systems. 
    We consider four variations of the lens mass model: 
    \begin{itemize}
        \item elliptical power law with free external shear used as a reference;
        \item elliptical power law with the external shear fixed to the best value from the previous case; 
        \item elliptical power law with 3rd order multipoles;
        \item elliptical power law with 3rd and 4th order multipoles.
    \end{itemize} 

    We model each variation of the dataset $\mathbf{\mathcal{D}_{1}}$  (Unperturbed, 1 keV, 3 keV) with each of these. In Tab.~\ref{tab:mock_123}, the priors are provided for all parameters. When a parameter is not included in the model, it is fixed at the fiducial value used to create the mock data. In the following, we will focus on this system as it is the most affected by the non-halo density perturbations. A complementary discussion, including the distributions for $x_{c}$, $y_{c}$, and $\lambda_{s}$ and the results for all three systems belonging to $\mathcal{D}_{1}$ is given in App.~\ref{app}. In the case of $\mathcal{D}_{2}$ and $\mathcal{D}_{3}$, we only fit the simple power law + external shear model to understand possible biases in $\mathcal{D}_{1}$.

\section{Results}\label{sec:results}

    \begin{figure*}
        \centering
        \subfloat[\centering Unperturbed]{{
        \includegraphics[width=0.9\textwidth]{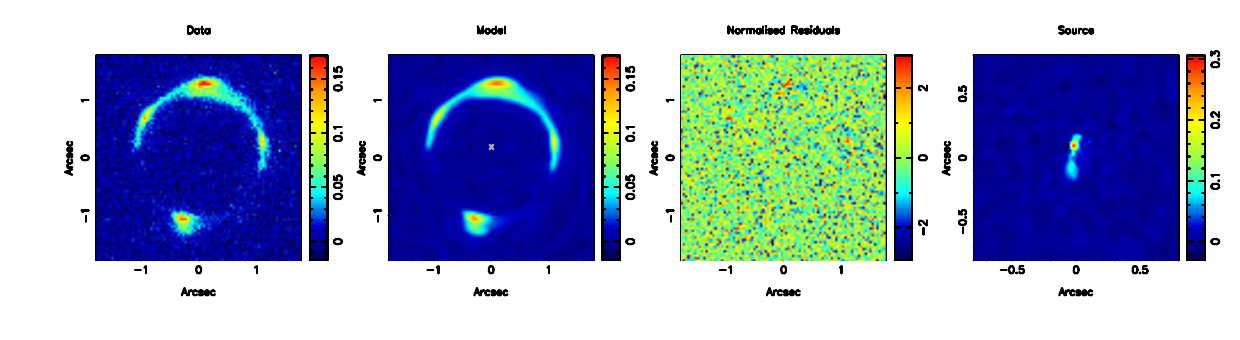} \vspace{-25pt}}}
        \qquad
        \subfloat[\centering 1 keV]{{
        \includegraphics[width=0.9\textwidth]{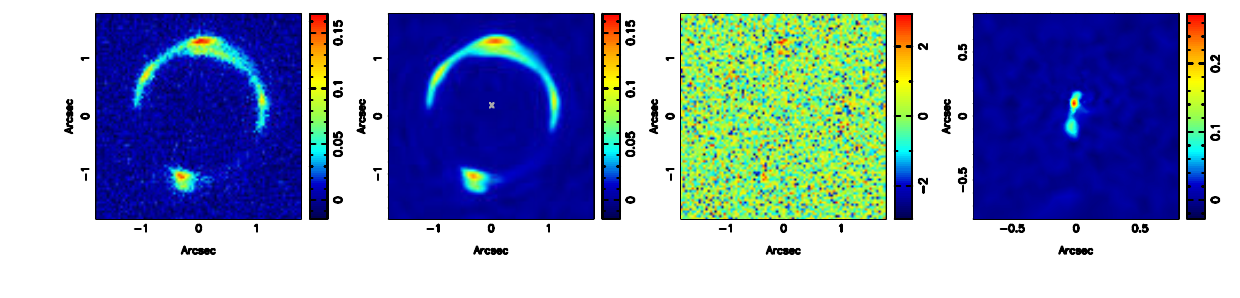} \vspace{-25pt}}}
        \qquad
        \subfloat[\centering 3 keV]{{
        \includegraphics[width=0.9\textwidth]{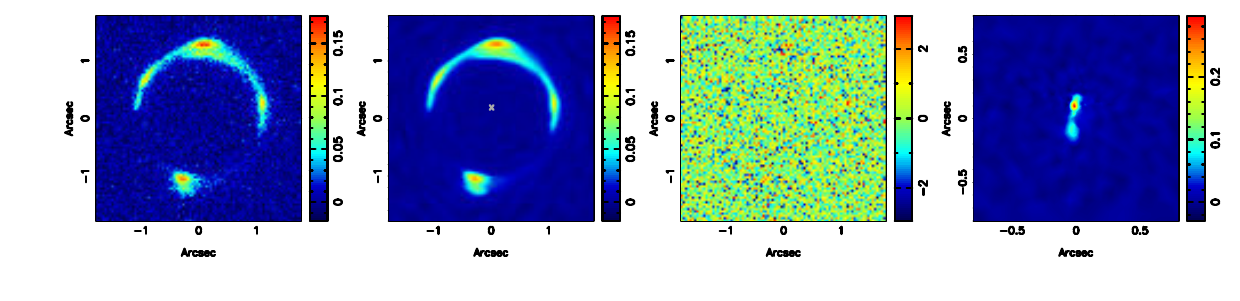} \vspace{-25pt}}}
        \caption{Optimisation results when modelling the three systems from $\mathcal{D}_{1}$ with an elliptical power law + external shear. The upper panel (a) shows the results for the Unperturbed system, the middle panel (b) for 1 keV, and the lower panel (c) for 3 keV. From left to right: images are the initial data, reconstructed model, normalised residuals between these two images, and reconstructed source. The units are arbitrary but are analogous to a 2D mass density}
    \label{fig:D1opti}
    \end{figure*}

    \begin{figure*}
        \centering
        \includegraphics[width=0.8\textwidth]{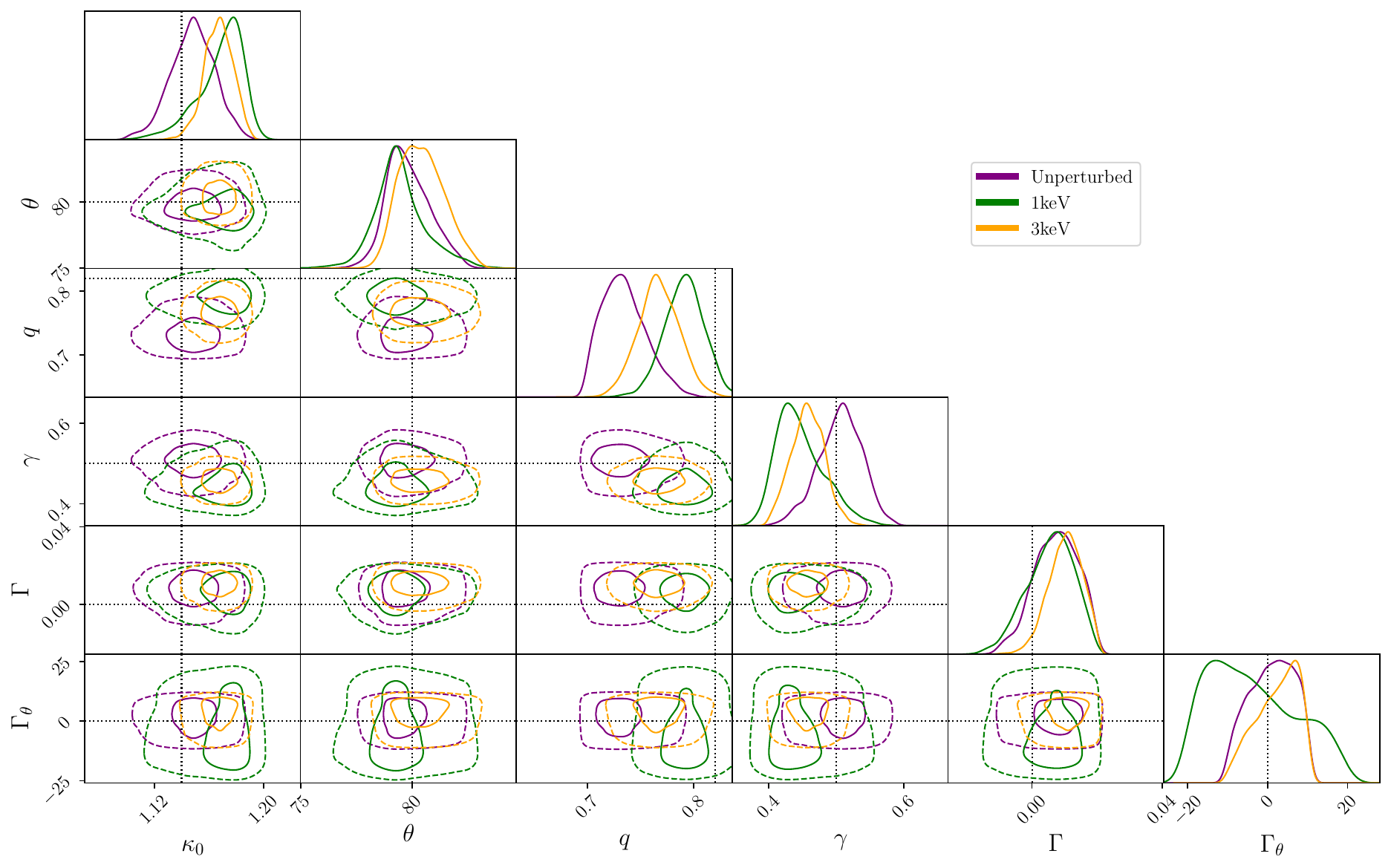}
        \caption{1$\sigma$ and 2$\sigma$ contours for the parameters $\boldsymbol{\eta_{m}}$ (excluding the center coordinates $x_{c}$ and $y_{c}$, and the source regularisation $\lambda_{s}$) for the elliptical power law + external shear model of the Unperturbed, 1 keV and 3 keV systems in $\mathcal{D}_{1}$. The dashed lines show the 2D distributions, the solid lines show the marginal distributions and the dotted grey lines mark the true values.}
    \label{fig:D1}
    \end{figure*}

    \begin{figure*}
        \centering
        \includegraphics[width=\hsize]{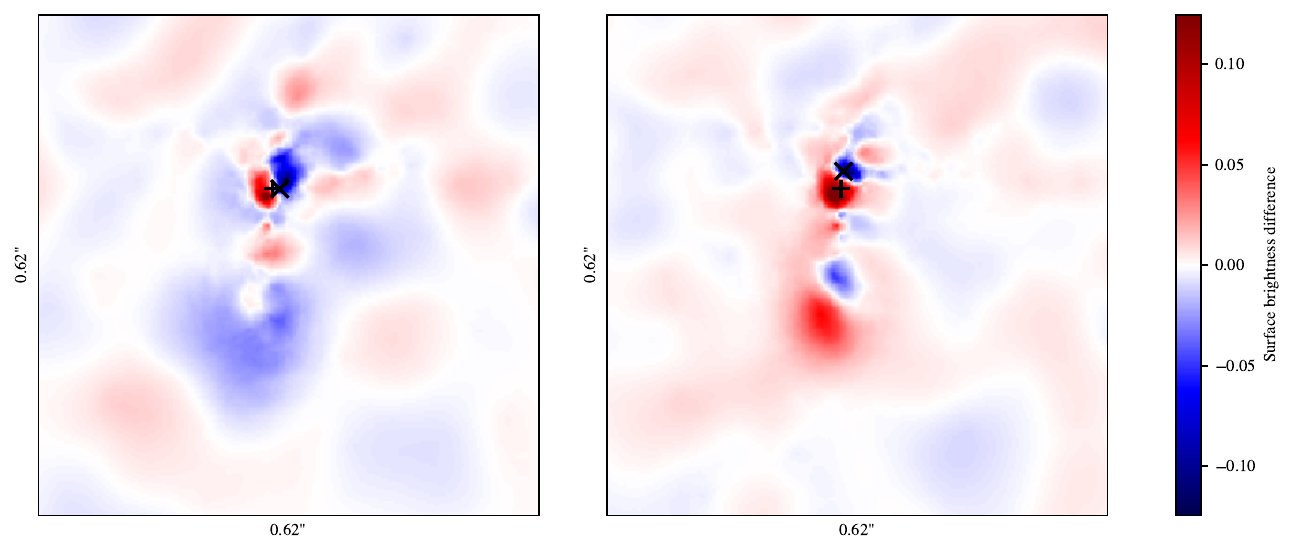}
        \caption{Difference in the sources brightness distribution in the source plane obtained when modelling the systems in $\mathcal{D}_{1}$ with a lens following an elliptical power law profile with additional external shear. If $S_{\text{X}}$ is the source surface brightness distribution obtained for the system X, these images are the result of $S_{\text{Unperturbed}}-S_{\text{1 keV}}$ (left) and $S_{\text{Unperturbed}}-S_{\text{3 keV}}$ (right). The local surface brightness of the reconstructed source can be shifted and modified in the perturbed models if some of the additional perturbations are absorbed by variations of the main lens properties. The general trend given by these images is that the main effect of the additional perturbation is an overall shift of the source, which is indicated by corresponding red and blue regions of the 2D brightness distribution. To highlight this effect, the black markers give the point of highest surface brightness ("+" for the unperturbed system, and "x" for the perturbed one).}
    \label{fig:D1diffs}
    \end{figure*}

    \begin{figure*}
        \centering
        \includegraphics[width=0.8\textwidth]{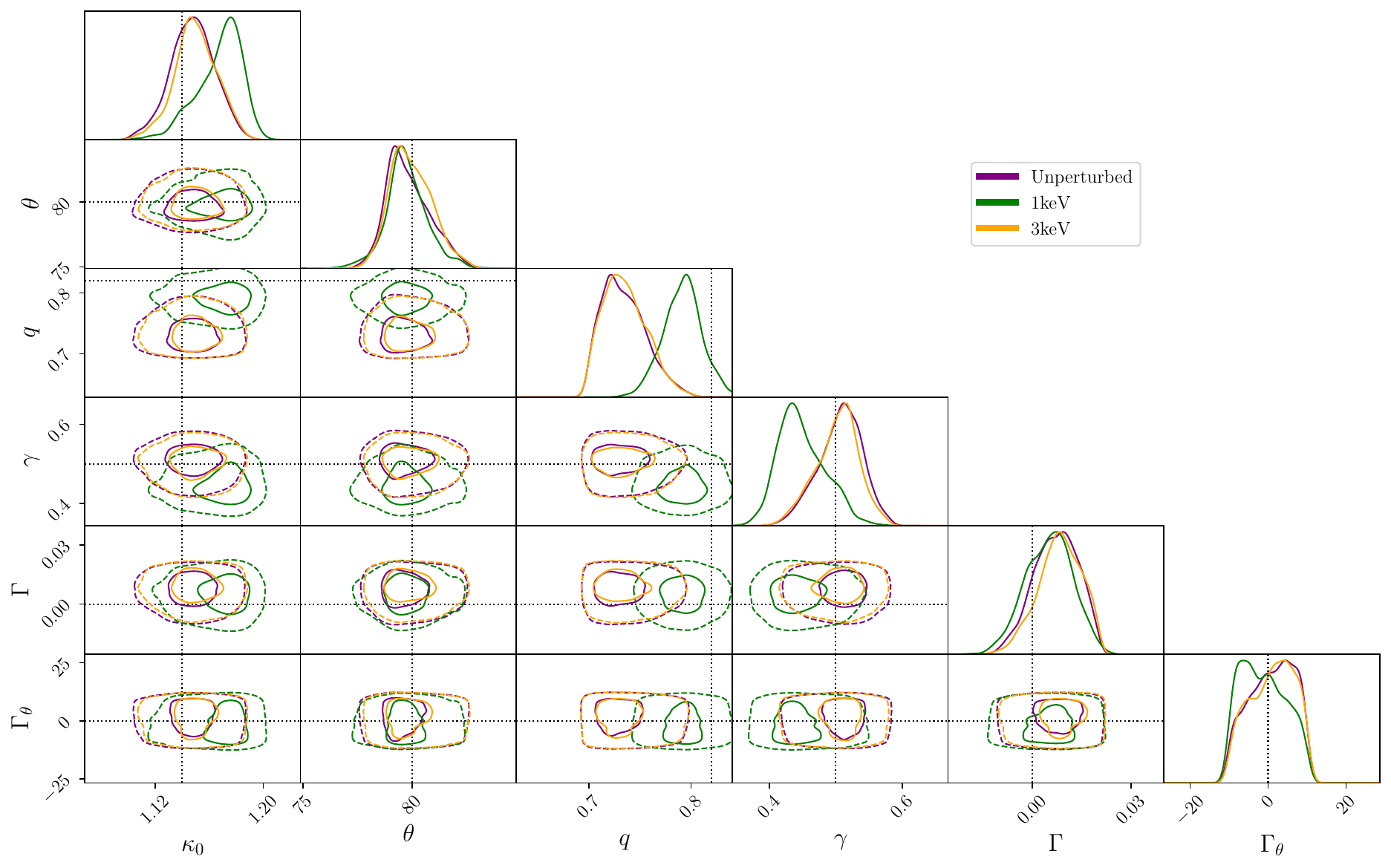}
        \caption{1$\sigma$ and 2$\sigma$ contours for the parameters $\boldsymbol{\eta_{m}}$ for the elliptical power law + external shear model of the Unperturbed, 1 keV and 3 keV systems in $\mathcal{D}_{2}$. The layout is the same as in Fig.~\ref{fig:D1}.}
    \label{fig:D2}
    \end{figure*}

    \begin{figure*}
        \centering
        \includegraphics[width=0.8\textwidth]{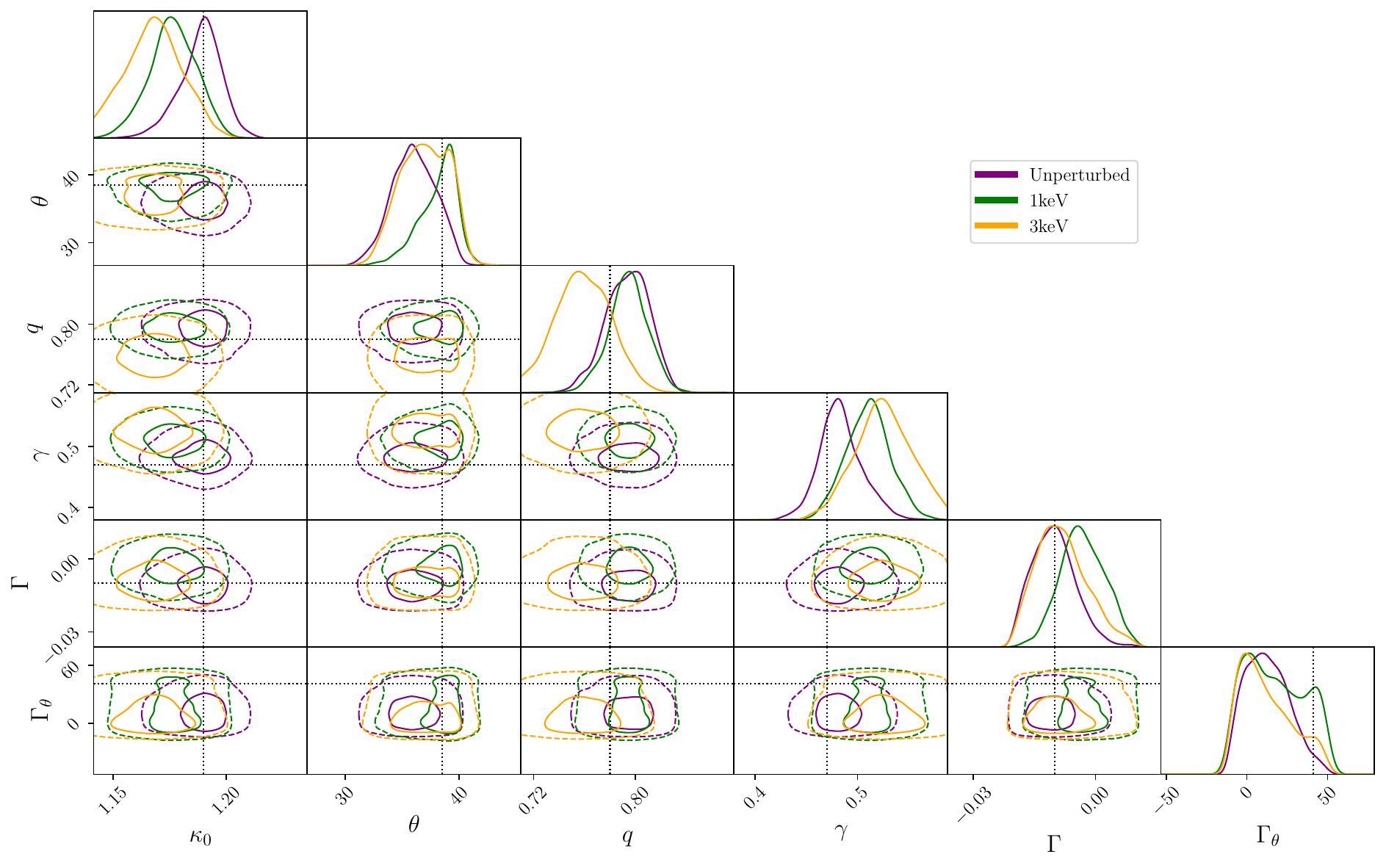}
        \caption{1$\sigma$ and 2$\sigma$ contours for the parameters $\boldsymbol{\eta_{m}}$ for the elliptical power law + external shear model of the Unperturbed, 1 keV and 3 keV systems in $\mathcal{D}_{3}$. The layout is the same as in Fig.~\ref{fig:D1}.}
    \label{fig:D3}
    \end{figure*}

    \begin{figure*}
        \centering
        \includegraphics[width=0.85\textwidth]{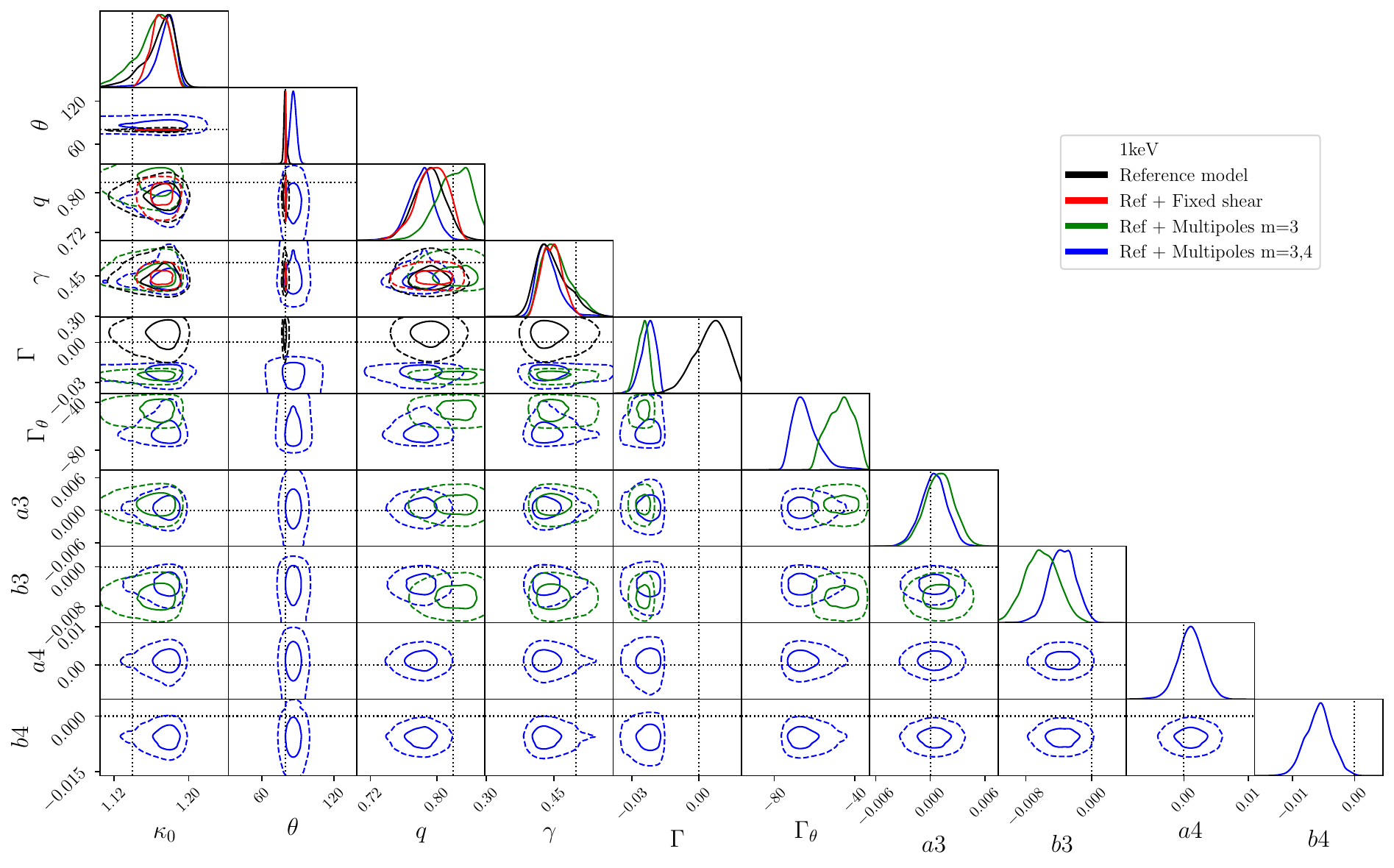}
        \caption{1$\sigma$ and 2$\sigma$ contours for the parameters $\boldsymbol{\eta_{m}}$ with third and fourth order multipoles (excluding the center coordinates $x_{c}$ and $y_{c}$, and the source regularisation $\lambda_{s}$) for the four models of the 1 keV system in $\mathcal{D}_{1}$. The reference lens model is the elliptical power law + external shear.}
        \label{fig:D11red}
    \end{figure*}

    In this section, we discuss the results of the lens modelling of mock images created with and without the inclusion of non-halo structures, as described in Sect.~\ref{sec:model}. For each dataset, we compare the evidence obtained by using the different mass models described in the previous section. The values of the difference in logarithmic evidence for each model are given in Tab.~\ref{tab:Ev_D1}.
    
    In this controlled experiment, we manually include the perturbations, as opposed to real observations where one does not know the underlying mass distribution of and around the main lens. We treat all the mock images as real observations and model them without imposing strong priors on the lens parameters. In this way, we want to test whether or not it is possible to distinguish the effect of the non-halo structures from that of the main lens. In particular, one could expect their effect to appear as (or increase) the external shear, as this term is expected to represent the effect of the environment around the lens. In fact, \cite{etherington2023strong} show that the external shear term compensates for the lack of model complexity rather than describing sources of large-scale shear. Our mocks allow us to test this hypothesis in a controlled setting.

\subsection{Are non-halo perturbations detected as external shear?}\label{ssec:shear}

    In Fig.~\ref{fig:maps}, we see that the shear amplitudes $\gamma_{1}$ and $\gamma_{1}$ are of the order of $10^{-3}$, which is similar to the range of values of external shear $\Gamma$ recovered from BELLS and SLACS lenses in \cite{Ritondale_2019} and \cite{etherington2023strong}. Therefore, we expect the effect of non-halo structures to be comparable with that of external shear. To examine this, we begin by modelling the mock observations of the $\mathcal{D}_{1}$ set with the simplest lens model, which includes only a power-law profile and external shear. In all cases, this simple model can fit the data quite well, leaving no persistent structures in the residuals, which have mean and median values close to zero, even in the presence of the non-halo convergence. Fig.~\ref{fig:D1opti} shows the three mocks (first column) and the results of the lens modelling. The corresponding posterior distributions are shown in Fig.~\ref{fig:D1}. If external shear describes the effect of large-scale structures, one would expect an increase (or variation) of the shear parameters in the perturbed case. Instead, we see that all lens parameters are affected, in particular the convergence normalisation $\kappa_{0}$, ellipticity $q$ and density slope $\gamma$. The posterior distribution for the shear strength $\Gamma$ and angle $\Gamma_{\theta}$ do not show a systematic difference compared to the unperturbed case, suggesting that the orientation of the filaments and walls does not leave a distinctive imprint. 
    
    Overall, the changes in the posterior distribution of the parameters are not significant, and the best-fit values for all parameters remain sensibly the same within the 1$\sigma$ error bars. The perturbations, no matter their orientations along the line of sight, seem to be reabsorbed by a reasonably smooth (unperturbed) model in both the 1 keV and 3 keV cases. Some of the differences in the mass model are absorbed by the source: as the source is pixelated and reconstructed at the same time as the lens model, the two are related, and the sources react to changes in the lens model. The sources in the right column of Fig.~\ref{fig:D1opti} overall show the same shape and visually look similar and in very good agreement with the modelling from \cite{Ritondale_2019}. However, Fig.~\ref{fig:D1diffs} shows that the differences between the resulting fine geometrical features and the source position shift lead to surface brightness differences up to half the maximum surface brightness of the source obtained for the Unperturbed system. As a check that the introduction of WDM perturbations does indeed shift the sources and does not alter the overall flux, we compute the ratio of integrated surface brightness $\frac{\Sigma S_{\text{1 keV}}}{\Sigma S_{\text{Unperturbed}}}$ and $\frac{\Sigma S_{\text{3 keV}}}{\Sigma S_{\text{Unperturbed}}}$ where the sums are performed over all the map pixels, and which respectively yield 0.97 and 0.98. Previous works \citep{brainerd2001gravitational, 10.1111/j.1365-2966.2005.09523.x, 10.1093/mnras/stt043} have shown that small variations in the source structure can absorb the perturbations due to low-mass perturbations and thus prevent their detection. In this case, we see that the source can also absorb part of the effect of the extended convergence caused by the non-halo large-scale structures given that it does not visibly alter the surface brightness of the arc in a localised way.

    We now model $\mathcal{D}_{2}$ and $\mathcal{D}_{3}$ to check if we get consistent results with the $\mathcal{D}_{1}$ analysis. Figs.~\ref{fig:D2} and~\ref{fig:D3} show the corresponding posterior distributions for these two datasets. We remind the reader that $\mathcal{D}_{2}$ is generated from the same base system as $\mathcal{D}_{1}$ but with perturbations rotated by +90°. This is a good sanity check that our results for $\mathcal{D}_{1}$ do not emerge for specific alignments between the perturbations and other lens or source features. 
    Comparing Figs.~\ref{fig:D1} and~\ref{fig:D2} gives us insights into the potential alignments between the source and the lens. 
    
    In Fig.~\ref{fig:D2} we observe that the recovered parameters lie in the same ranges as Fig.~\ref{fig:D1}, where in both cases we observe significant overlap between the posteriors corresponding to the unperturbed and perturbed lenses. 
    This is a good qualitative indication that no parameter is strongly affected by the orientation of the perturbation. However, we can appreciate some small differences: for $\mathcal{D}_{2}$, the parameters remain the same between the Unperturbed and the 3 keV systems and vary only when introducing the 1 keV perturbations, which are stronger. 
    In $\mathcal{D}_{1}$, we instead see a gradual shift of the posterior distribution for $\kappa_{0}$, $q$ and $\gamma$ when perturbations are introduced.
    Given that the perturbation orientation is the only difference between the two cases, this indicates that in $\mathcal{D}_{2}$ the alignment with the lens can conspire to mask their effect - reabsorbing it into the model even more - in the 3 keV case, as hinted by the convergence map in Fig.~\ref{fig:maps}.

    In $\mathcal{D}_3$ (Fig.~\ref{fig:D3}), both the lens model and the input source are different. We nonetheless observe that the non-halo structures have a similar impact on the posteriors than that already discussed for $\mathcal{D}_1$ and $\mathcal{D}_2$. 
    This confirms that different perturbations alter the lenses, but that they remain in statistically good agreement with each other as all the 1$\sigma$ contours overlap.

\subsection{Effect of lens model complexity}\label{ssec:comp}

    In the previous section, we have shown that an elliptical power-law + external shear parametrisation provides a good model to reconstruct the mock observations. In this Sect., we explore how non-halo structures affect more complex models and if this increased complexity can better capture these perturbations. As such, we analyse each mock observation within the $\mathcal{D}_{1}$ dataset with four different mass models and can compare the resulting $\Delta$log($\mathcal{E}$) values displayed in Tab.~\ref{tab:Ev_D1}. As a first test, we fix the external shear parameters to the best fit of the power-law + shear model. This reduces the number of parameters but decreases the quality of the lens modelling. The discrepancy between the evidence values gets bigger with the addition of non-halo perturbations, as $\Gamma$, even when it is consistent with 0 at the 1$\sigma$ level, can absorb some of these perturbation effects. We note that when $\Gamma$ is consistent with 0, the posterior distribution for the parameter $\Gamma_{\theta}$ does not converge as one cannot assign a meaningful direction to the sheer in this case (see Figures~\ref{fig:D1}, ~\ref{fig:D2}, ~\ref{fig:D3}, ~\ref{fig:D11red}, ~\ref{fig:D10},~\ref{fig:D11} and~\ref{fig:D13}).

    We then consider a mass model with additional complexity given by third and fourth-order multipoles. While the input mass model used to create the mock image does not explicitly contain these multipoles one may expect that they may be able to capture additional features originating from the non-halo convergence. We find that this is however not the case as for all three datasets (Unperturbed, 1 keV, 3 keV), the addition of multipoles (m=3 or m=3,4) to the model decreases the evidence. Indeed, we observe that the blue and green distributions of Fig.~\ref{fig:D11red} shift away from the black and red ones, corresponding more to the input source, and are also broader, hence giving less precise constraints on the values of the parameters. Overall, the models with multipoles are less preferred and the posterior distributions of the parameters are less constraining, and do not help capture the effect of the perturbations. Nonetheless, it is important to note that all multipoles parameters are consistent with 0 at the 1$\sigma$ level. Looking at the differences between the values of $\Delta$log($\mathcal{E}$) for all three systems, it could be tempting to assume that since they are lower in the 1 keV scenario, multipoles are better at absorbing 1 keV non-halo features than their 3 keV counterparts. However, one should not compare these three systems with each other, and we remind the reader that the perturbation introduced with the 1 keV and 3 keV simulations display different geometrical features. 

    Furthermore, looking at the distributions in Fig.~\ref{fig:D11red}, we can notice several similarities and differences between our four models for the 1 keV system (that are similar for the Unperturbed and 3 keV systems). Most parameters follow similar posterior distributions in different models within 1$\sigma$. For all the systems, it appears that every parameter's posterior distribution converges. The Bayesian modelling used here \citep{10.1111/j.1365-2966.2008.14005.x} can provide optimised values for the lens profile in the four different types of models tested, and the presence of the perturbations is marked by minor shifts in the posterior distributions of the parameters. Thus, the perturbations introduced in the strong lensing signal by WDM non-halo structures appear as a systematic effect and are not equivalent to multipoles.

\section{Conclusions}\label{sec:conc}

    In this study, we have investigated if and to what extent the material outside of haloes, i.e. within filaments and walls, can cause statistically relevant and observable effects on gravitationally lensed arcs in the case of galaxy-galaxy lensing. Such effects could be used to constrain the dark matter particle mass or, equivalently, warmth. Here, we tested the effects of non-halo material from two simulations of WDM cosmological scenarios with dark matter particle masses $m_{\chi}$ = 1 keV and $m_{\chi}$ = 3 keV. So far, most studies have focused on the effects of dark matter haloes on gravitationally lensed systems, but a recent study (\cite{10.1093}, R22 in this work) has shown that non-halo structures have an effect on quadruply lensed quasar flux ratio observations.
    Despite numerous simplifications, the authors show that in a $m_{\chi} = 3$ keV scenario, neglecting the non-halo material can lead to an underestimation of the flux ratios by $5$ to $10$ per cent, while this number goes up to $50$ per cent for $m_{\chi} = 1$ keV. Although this dark matter model is already observationally ruled out, the authors conclude that non-halo material should be included in rigorous models and that their inclusion can lower the constraints on the dark matter particle mass, allowing for the more permissive study of cosmologies.

    Here, we have used the same WDM hybrid sheet and release simulations with $m_{\chi} = 3$ keV and $m_{\chi} = 1$ keV as in R22. These are novel fragmentation-free simulations, where particles are separated into four classes: voids, pancakes, filaments, and halos. Three of these classes, voids, pancakes and filaments, are simulated using $256^{3}$ sheet tracing particles, while the fourth class, haloes, are simulated using $512^{3}$ $N$-Body particles. The sheet interpolation scheme used for the first three classes allows for the resampling of the density field in these regions with unprecedented mass resolution. From the simulated density fields, we obtain non-halo structure convergence maps that we add as perturbations to two gravitationally lensed systems from the BELLS-Gallery.
    We analyse the resulting mock observations using four different models for the lens mass distribution of the systems in the first dataset and and with only one of these models in the second and third dataset. 
    Classic lens models describe the main lens with a single elliptical power-law (with or without additional multipoles), while the contribution of matter on larger scales is represented by an external shear term. Our simulated observations allow us to test whether or not this term actually corresponds to extended mass components, such as filaments and walls, and if they can be distinguished from the main lens model. Our main conclusions are:

    \begin{itemize}
    \item Of all tested models, we find that a single elliptical power-law deflector with additional external shear provides the best fit. Removing external shear or adding multipoles does not help to account for the potential effects of non-halo structures. In fact, all multipole parameters are consistent with 0 at the 1$\sigma$ level.
    \item When jointly reconstructing the lens and source, WDM perturbations are absorbed by the model parameters but introduce a systematic bias into both the source surface brightness distribution and the lens mass-density distribution. The main effect on the reconstructed source is small localised displacements with no apparent global modulation of the shape and surface brightness. 
    \item We find that the commonly used singular elliptical power-law lens + external shear parametrisation is sufficient to account for most of the effect at the cost that these perturbers cannot be distinguished from others that are also absorbed by this parametrisation \citep{Despali_2021}.
    \end{itemize}

    We conclude that in the case of galaxy-galaxy lensing, the effects of filaments and walls, i.e. material outside of haloes, can impact the reconstruction of the source surface brightness in a non-trivial way when using common lens mass profiles. However, this effect remains negligible in WDM cosmologies with $m_{\chi}$=1 keV and $m_{\chi}$=3 keV, and therefore for colder cosmologies too, where the relative importance of non-haloes compared to the one of haloes is lower.
    
    The simulated WDM non-halo material studied in this work originates from the same simulations as in R22, thus we have to take the same precautions as them when interpreting our results, and our limitations are of similar natures: we used the thin-lens approximation, baryonic effects are not modelled, dark matter particles of such warmth are already observationally excluded, and the line of sight is relatively short (80$h^{-1}$Mpc) compared to typical gravitationally lensed systems sizes. Thus, it is not possible to thoroughly quantify the effect of the non-halo structures of the line of sight, and we only consider their relative effects in the region of the main lens. Placing all non-halo structures in the lens plane maximizes their effect because lensing strength scales with convergence, which is highest at that redshift. In reality, mass distributed along the line of sight contributes less due to geometric dilution and shear cancellations. Additionally, multi-plane deflections introduce non-linearities that tend to weaken the perturbations compared to a single-plane approximation. Thus, our approach provides an upper bound on the effect — if it is small here, it would be even smaller in a full ray-tracing treatment.

    It is interesting to note the apparent mismatch between the conclusions of this work and those of R22. Indeed, R22 find that non-halo structures introduce a 5–10 per cent perturbation to flux ratio anomalies, our analysis shows that their effects on extended, pixellated sources are largely absorbed into the lens model. This difference arises because flux ratios are sensitive to magnification perturbations (convergence and shear), whereas extended source reconstructions primarily respond to deflection field perturbations. The latter are generally less affected, as also seen in R22’s appendix C, where non-halo structures more efficiently perturb brightness than image positions. This is due to the intrinsic difference in the data and the lens modelling process: a point source in the flux ratio analysis cannot absorb surface brightness variations as the pixellated source considered here. Moreover, the flux ratio analysis probes all unaccounted mass in the lens or along the line-of-sight, while in the modelling of lensed arc, one tries to identify which component is causing an extended surface brightness variation. This distinction explains why non-halo effects appear more significant in flux ratio studies but remain negligible in our modelling approach.

\begin{acknowledgements}

      We acknowledge helpful comments and corrections from the anonymous referee. We acknowledge insightful comments on the manuscript and direction for this project from Simona Vegetti.
      B. J. is supported by a CDSN doctoral studentship through the ENS Paris-Saclay and by the Max Planck Institute for Astrophysics. G. D. acknowledges the funding by the European Union - NextGenerationEU, in the framework of the HPC project – “National Centre for HPC, Big Data and Quantum Computing” (PNRR - M4C2 - I1.4 - CN00000013 – CUP J33C22001170001). T. R. acknowledges funding from the Spanish Government's grant program ``Proyectos de Generaci\'on de Conocimiento'' under grant number PID2021-128338NB-I00. J. S. acknowledges support from the Austrian Science Fund (FWF) under the ESPRIT project number ESP 705-N.
      We made extensive use of the {\tt numpy} \citep{oliphant2006guide, van2011numpy}, {\tt scipy} \citep{2020SciPy-NMeth}, {\tt astropy} \citep{1307.6212, 1801.02634}, {\tt GetDist} \citep{2019arXiv191013970L}, and {\tt matplotlib} \citep{Hunter:2007} python packages.
      The data used and produced in this work is available upon reasonable request to the corresponding author.
      
\end{acknowledgements}

\bibliographystyle{aa}
\bibliography{main}

\onecolumn
\begin{appendix}

\section{Details on the system modelling}\label{app}

\begin{figure}[h!]
        \centering
        \includegraphics[width=0.9\textwidth]{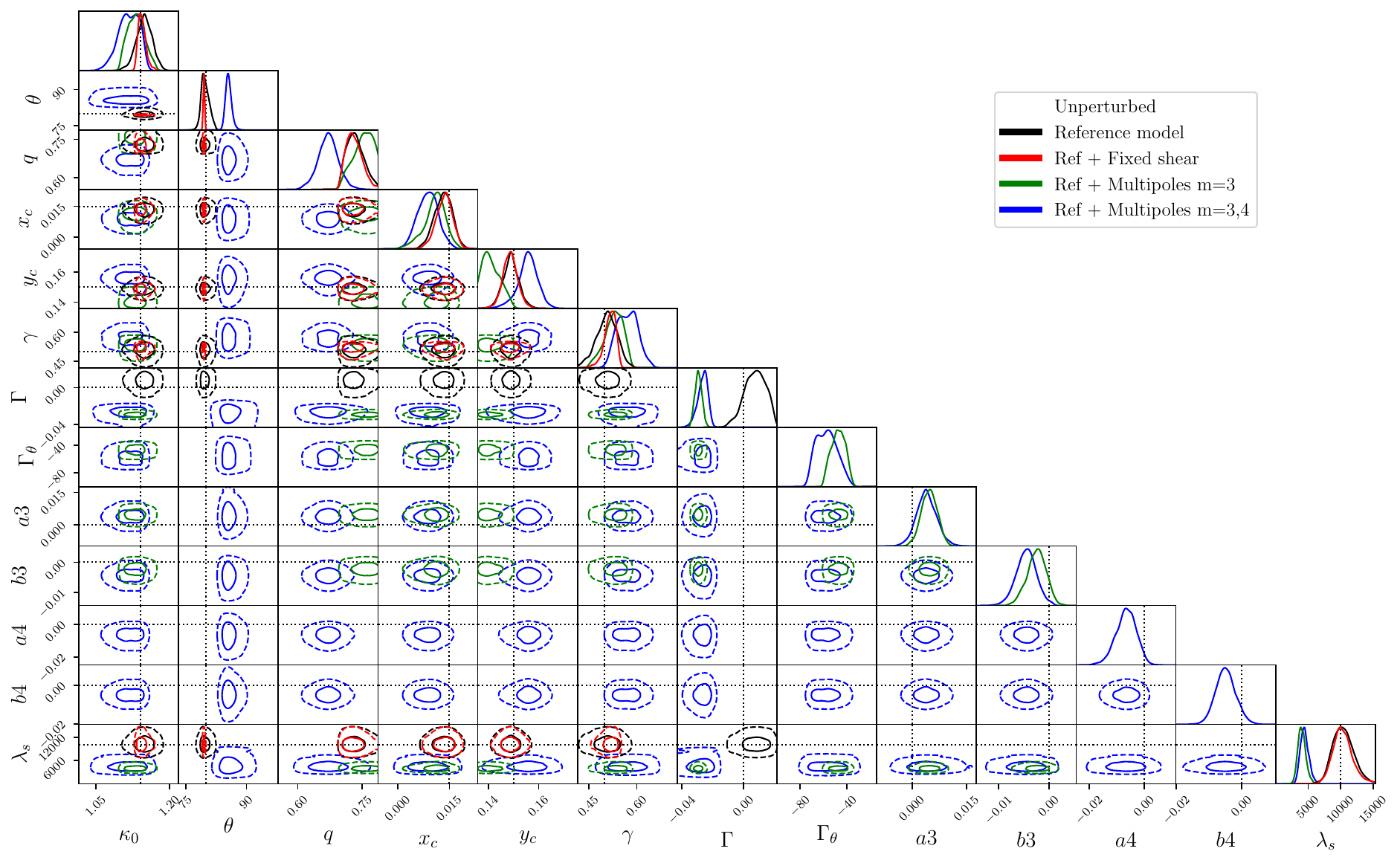}
        \caption{1$\sigma$ and 2$\sigma$ contours for the parameters $\boldsymbol{\eta_{m}}$ and $\lambda_{s}$ for the four models of the Unperturbed system in $\mathcal{D}_{1}$.}
        \label{fig:D10}
    \end{figure}

\begin{figure}
        \centering
        \includegraphics[width=0.9\textwidth]{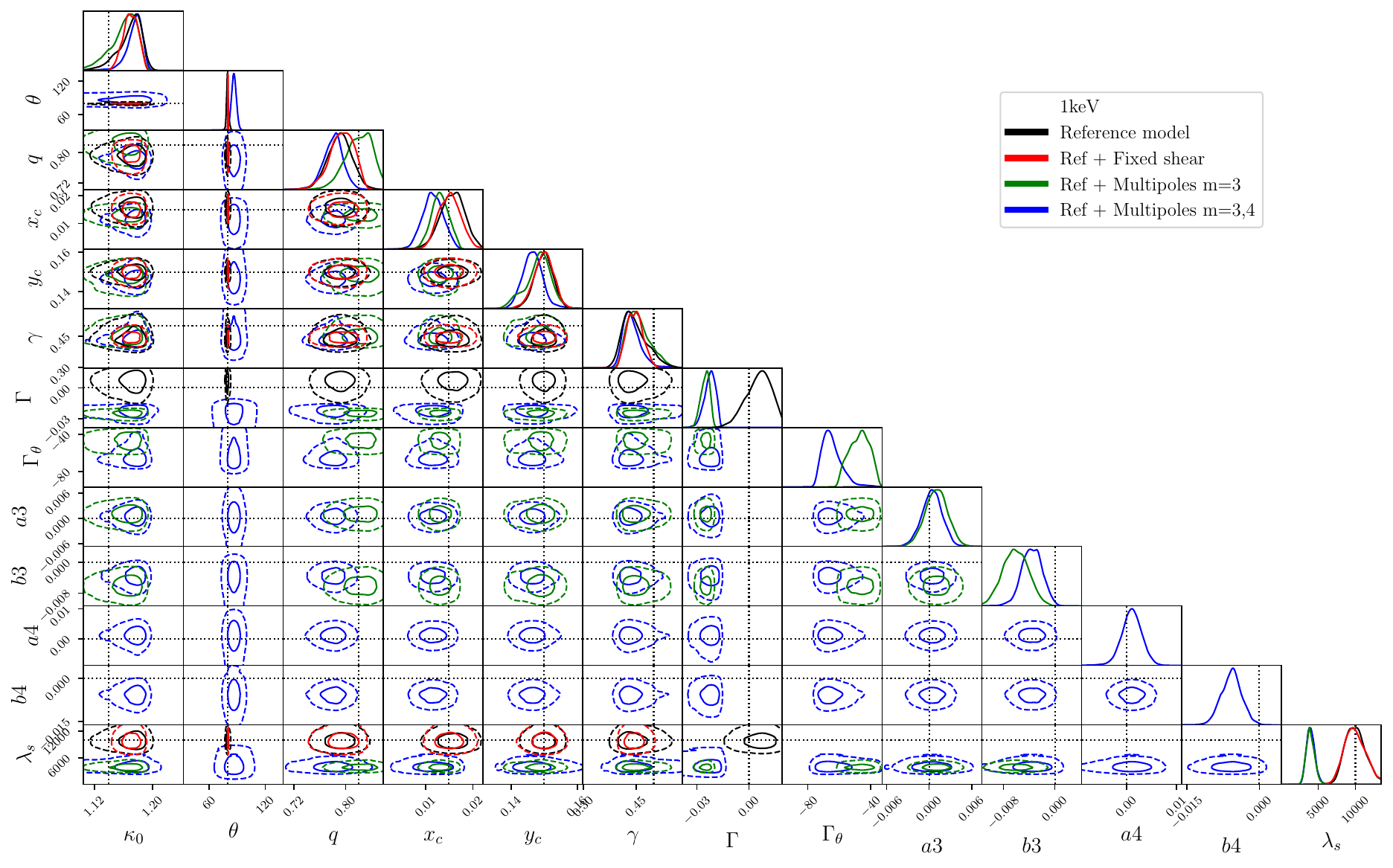}
        \caption{1$\sigma$ and 2$\sigma$ contours for the parameters $\boldsymbol{\eta_{m}}$ and $\lambda_{s}$ for the four models of the 1 keV system in $\mathcal{D}_{1}$.}
        \label{fig:D11}
    \end{figure}

\begin{figure}
        \centering
        \includegraphics[width=0.9\textwidth]{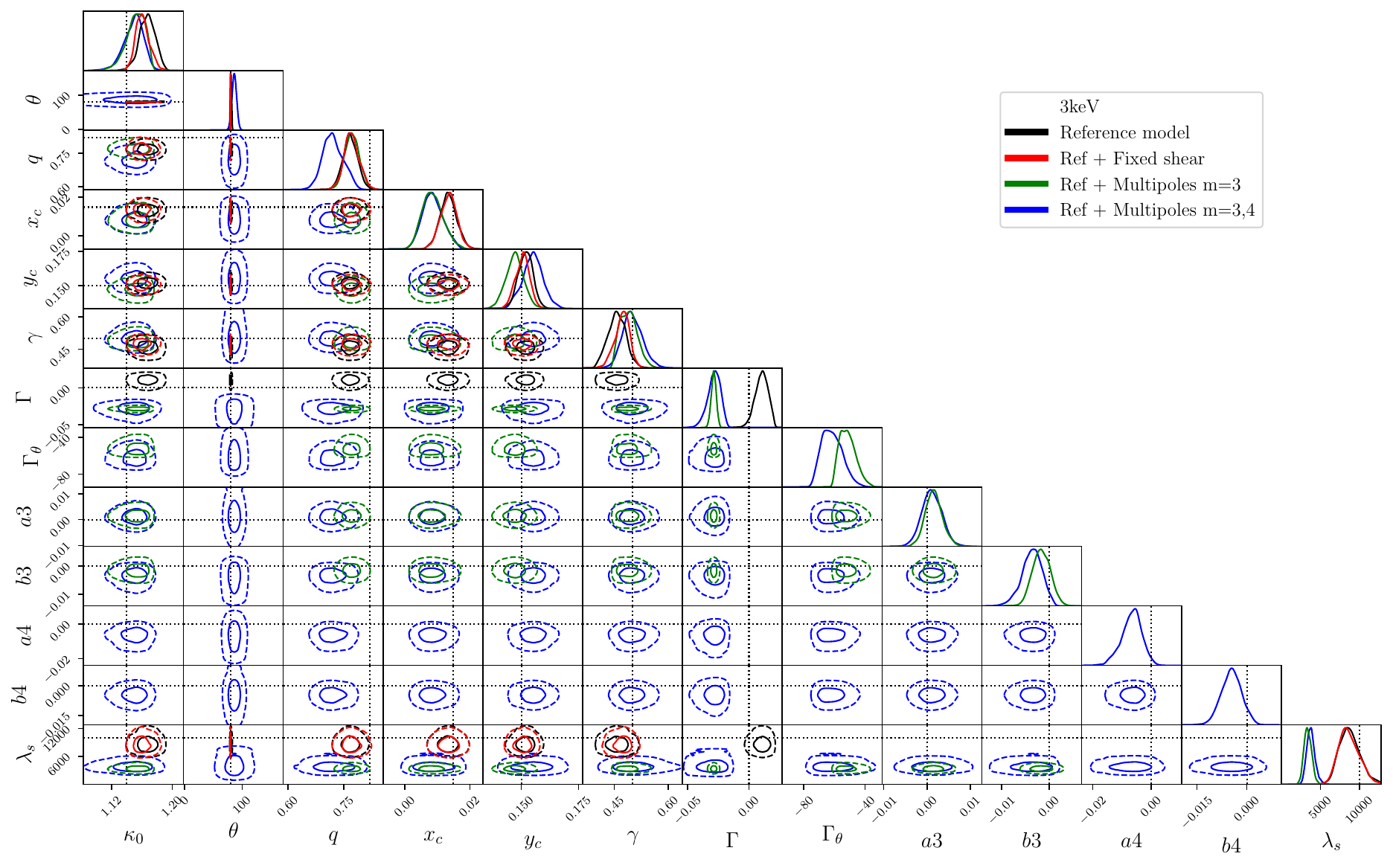}
        \caption{1$\sigma$ and 2$\sigma$ contours for the parameters $\boldsymbol{\eta_{m}}$ and $\lambda_{s}$ for the four models of the 3 keV system in $\mathcal{D}_{1}$.}
        \label{fig:D13}
    \end{figure}

    Figures~\ref{fig:D10},~\ref{fig:D11} and~\ref{fig:D13} give the posterior distributions for all the lens parameters used to model the three systems in $\mathcal{D}_{1}$.

\end{appendix}

\end{document}